\let\ifpdf\relax
\documentclass{JINST}
\let\ifpdf\relax

\title{Non-planar four-mirror optical cavity for high intensity gamma ray flux
 production by pulsed laser beam Compton scattering off GeV-electrons}

\author{J. Bonis$^a$, R. Chiche$^a$, R. Cizeron$^a$, M. Cohen$^a$, E. Cormier$^b$, P. Cornebise$^a$, N. Delerue$^a$\thanks{Corresponding
author.}, R. Flaminio$^c$, D. Jehanno$^a$, F. Labaye$^a$, M. Lacroix$^a$, R. Marie$^a$, B. Mercier$^a$, C. Michel$^c$, Y. Peinaud$^a$, L. Pinard$^c$, C. Prevost$^a$, V. Soskov$^a$, A. Variola$^a$ and  F. Zomer$^a$\\
\llap{$^a$}Laboratoire de l'Acc\'el\'erateur Lin\'eaire, CNRS-IN2P3 Universit\'e Paris-sud 11 ,\\
15, rue Cl\'emenceau, F-91898 Orsay Cedex, France\\
\llap{$^b$}CELIA, CNRS, Universit\'e Bordeaux 1 ,\\
  43 rue Pierre Noailles, F-33405 Talence, France\\
\llap{$^c$}LMA, CNRS-IN2P3 Universit\'e Lyon 1 ,\\
7, Avenue Pierre de Coubertin, F-69622 Villeurbanne Cedex, France\\
  E-mail: \email{delerue@lal.in2p3.fr}}

\abstract{
As part of the R\&D toward the production of high flux of polarised Gamma-rays we have designed and built a non-planar four-mirror optical cavity with a high finesse and operated it at a particle accelerator. We report on the main challenges of such cavity, such as the design of a suitable laser based on fiber technology, the mechanical difficulties of having a high tunability and a high mechanical stability in an accelerator environment and the active stabilization of such cavity by implementing a double feedback loop in a FPGA.
}


\begin{document}

\section{Introduction}
Recent results in cultural heritage conservation~\cite{louvre}, medecine~\cite{med1,med2,med3}, nuclear~\cite{nuphy1,nuphy2}  and particle physics~\cite{araki,goodie,gamma_gamma} have shown that there is a strong interest in producing high X and $\gamma$ ray fluxes by Compton scattering  a laser beam onto an electron beam. 

Two ambitious projects are starting \cite{quantumbeam,ThomX} to provide a compact monochromatic X-ray source \cite{compactref} from  the scattering of a laser beam on electrons with an energy in the $10-100$MeV range for medical imagery and heritage applications. Because of the small Compton cross section, a laser beam average power between 100kW and 1MW is needed. 

At high energy, in the context of the linear collider projects CLIC\cite{CLIC} and ILC\cite{ILC},
 the gamma rays produced by Compton scattering of a circularly polarized laser beam on a few GeV electron beam can be used to create a longitudinally polarized positron beam \cite{araki,goodie,CLICpositrons} or high energy gamma beams \cite{gamma_gamma}. Nevertheless, this application of Compton scattering requires a huge laser beam average power above the Megawatt level. 

The technology to reach the extremely high average laser power required for these applications is not yet mature. Direct amplification technology is highly inefficient and does not stands large average power (see e.g. \cite{ref_ampli}). Instead the use of an optical resonator \cite{kogel}, {\it i.e.} a Fabry-Perot cavity (FPC),  of very high finesse filled with a pulsed laser beam \cite{ruth,nestor} allows to reach a high power at the collision point with easier requirements on the amplification and a much higher efficiency. If the electron beam passes inside the FPC one can indeed benefit from the power gain, defined by the finesse over $\pi$,  of the resonator provided that resonance conditions are fulfilled. However, a strong feedback is needed to lock a mode lock laser beam to a FPC \cite{feedback_modelock}. In optical laboratories, a stored average power of  $\simeq$70kW was achieved recently \cite{les70kw} and the operation of cavity finesse of  $\simeq$30000 was demonstrated \cite{nousnim}. In accelerator environments very few experiments have been carried out. In continuous regime, a cavity finesse of  $\simeq$30000 has been operated routinely \cite{jlab,HERAnous} and in the pulsed regime cavities of finesses $\simeq$3000 on a short linear accelerator \cite{sakaue} and  $\simeq$1000 on the ATF ring \cite{atf2m1} have been operated. 

In order to  increase the Compton $\gamma$-ray flux one must not only increase the stored laser pulse power but also reduce the waist of the cavity mode.  It is well known \cite{kogel} that two-mirror cavities become unstable when the mode waist decreases and that four mirrors, or more, must be considered in this case (see \cite{nous4mirroir} for example). All resonators having being operated in accelerators up to now were made of two mirrors. 

In this article, we describe a new experimental apparatus installed at the KEK ATF~\cite{ATF} that will contribute to a global R\&D effort to reach the 100kW level of stored average power inside a four-mirror cavity. A new, non-planar geometry is used for the first time to provide highly stable and circularly polarized cavity eigenmodes~\cite{nous4mirroir}. The results reported here are those obtained with our apparatus before the earthquake that struck Japan in March~2011 and therefore we expect to be able to improve them once the apparatus will have fully recovered and be ready for operations again. 

\begin{figure}[htbp]
\begin{center}
\hspace*{-2cm}\includegraphics[width=.9\textwidth]{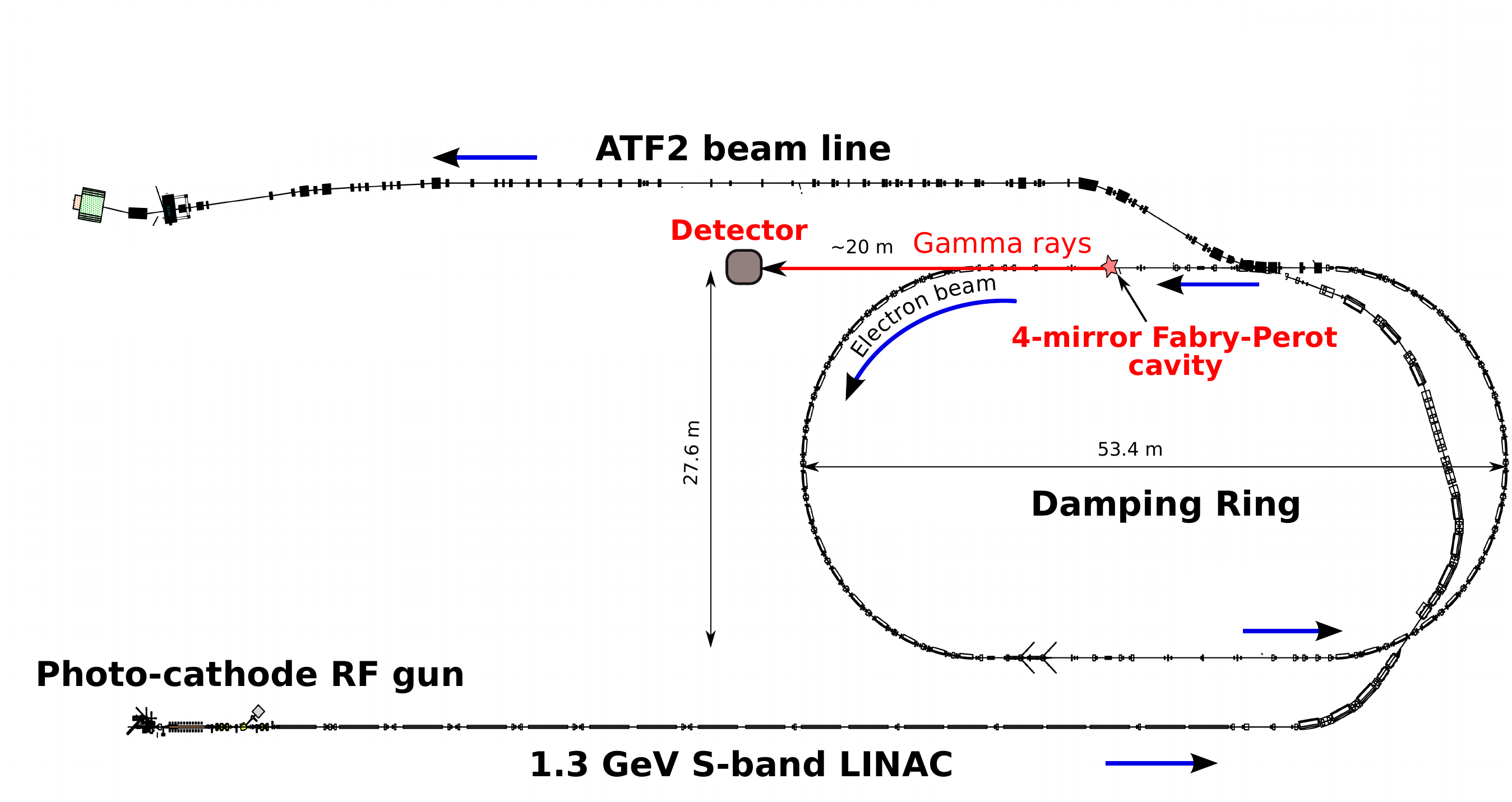}
\caption{The Accelerator Test Facility (ATF) at KEK. The green box indicates the approximate locate of  the 4-mirror cavity described in this paper. The red arrow indicates the direction of travel of the  gamma ray produced. The blue box indicates the gamma-ray detector (adapted from~\cite{ATF2}).}
\label{fig:ATF_DR}
\end{center}
\end{figure}

The ATF damping ring at KEK~\cite{ATF,ATF2} operates at a frequency of 357MHz and a round trip in the ring takes 462ns (there are 165 RF buckets spaced by 2.8ns). Up to 3 trains of 10 bunches separated by 5.6ns can be injected in the ring however most operations are done with a single bunch in the ring.  Our FPC is installed in one of the straight sections of the damping ring as shown on figure~\ref{fig:ATF_DR}.

The laser oscillator that can be locked to a very high finesse cavity must exhibit the smallest possible phase (or frequency) noise.
Such oscillator deliver typically only an average power of about 100mW. Even with a cavity finesse of  $\simeq$30000 and a perfect laser beam/cavity mode coupling  one could only reach $\simeq$1kW average power inside the cavity.  A laser amplification stage, designed to reach high average power without contributing too much additional frequency noise,  is therefore also needed to meet the performance requested for the applications mentioned above. In addition, the laser beam must be circularly polarized for the polarized positrons source application. 

To reach a high incident average laser power, we choose to use an  Yb doped photonic fiber chirp pulse amplifier. This recent technology is very promising, at present $\approx$800W average power were obtained with a 78~MHZ repetition rate, 640~fs pulse time width laser beam \cite{800W}. The long term stability and reliability are however still crucial issues and remain to be demonstrated. In this article, we report, for the first time to our knowledge,  on the operation of a 50W amplifier installed close to the FPC in the ATF accelerator.

 The mechanical design of the cavity includes a Compton scattering interaction point and fulfills the strong requirements set by the ATF accelerator. These stringent requirements lead us to develop new high precision mirror mounts for ultra high vacuum.

We implemented the Pound-Drever-Hall (PDH) laser/cavity feedback method \cite{PDH} using a fully numerical system based on a field programmable gate array (FPGA). FPGAs have already been used in a recent past to lock laser oscillators to reference frequencies. 
  In addition, the relatively  easy programming of the feedback filters and integrators offers a flexibility that helps to perform the complex locking of a laser frequency comb \cite{udem_nature} to a cavity \cite{locking_jones,locking_je}.
  However, an analog part is still often included either in the feedback system or  for the measurement of the response functions required to tune the feedback. 
 The system presented in this article contains, for the first time to the authors' knowledge, all the feedback bricks inside the FPGA, that is the frequency modulation/demodulation, the error signal filtering, the production of the correction signals and the identification procedure leading to the response functions.

The commissioning of our apparatus at the ATF has been successfully done during winter 2010-2011. Since many new experimental features have been developed to build the experimental setup, the present article is solely devoted to the apparatus. The first results obtained during the commissioning, before the earthquake of March 2011, are published in a companion article~\cite{analysis_Iryna}. 

This article is organized as follows. In section \ref{section_opto} the optical system , the laser amplification and the FPC are  described. The mechanical system  is discussed in  \ref{section_meca}. The electronic and the feedback system are presented in section~\ref{section_electronique}.

\section{The Optical system} \label{section_opto}

\subsection{Requirements}





To deliver a beam suitable for injection in the FPC the seed laser oscillator 
must provide 
a frequency comb in frequency space~\cite{udem_nature} with an
extremely stable repetition rate $f_{rep}$ and Carrier-Envelop Phase
(CEP). The comb tooths are defined by \cite{udem_nature,feedback_modelock}:
$$
\nu_n=\left( n+\frac{\Delta(\phi_{ce})}{2\pi} \right) f_{rep}
$$
where $f_{rep}$ is the laser pulse repetition rate (178.5MHz in our case); $n\approx 10^6$ is an integer number; $\Delta(\phi_{ce})$ is the carrier-envelope phase variation between two successive laser pulses.  From this expression one can sees that to lock a full frequency comb to a FPC, one must control two different parameters 
  $f_{rep}$ and $\Delta(\phi_{ce})$. It was shown that the PDH method \cite{PDH,PDH2} can provide  error signals as requested to perform the laser beam / FPC synchronisation~\cite{locking_jones,locking_je}. In practice the dispersion induced by the propagation inside the cavity mirror coatings limits the number of combs that can be locked to a FPC and introduces some coupling losses \cite{peterson}. In our setup, the width of the incident laser beam spectrum is reduced to 2~nm (see section \ref{section-opto}). We computed the variation of the cavity power coupling as a function of $\Delta(\phi_{ce})$. Assuming that $f_{rep}$ is locked to a FPC  round trip and a finesse of 3000, we obtained an expected maximum power coupling loss of $\approx 35\%$ for $\Delta(\phi_{ce})\in [0,2\pi]$ and $\approx 90\%$ for a 30000 FPC finesse. In the present article, a 3000 cavity finesse is used and no attempt was made to control  $\Delta(\phi_{ce})$. 

Another important feature of the FPC behavior in pulsed regime is the longitudinal mode structure. When the FPC length is detuned by $m\lambda$, where $\lambda$ is the laser beam wavelength, secondary resonances with lower power coupling are observed \cite{feedback_modelock}. In \cite{these_slac} it was shown that an effective cavity bandwidth can be defined and that, for a longitudinal resonance of order $m$, it increases with $m$. In our experiment we made use of this property by locking our FPC to the $m=1$ resonance during the very first step of the commissioning at ATF before switching to the main resonance ($m=0$).

As the power required to inject the passive FPC
exceeds by far the output power of any low phase noise oscillator, the
seed beam must be amplified in an optical amplifier. Here again, a
special attention should be taken in designing the amplifier system as
it should not add any phase noise to the seed beam in order to
maintain a high coupling in the FPC and reach the theoretical
FPC gain.
 The spatial quality of the amplified beam is also important to ensure 
an efficient coupling. Ideally  it should optimally match the
nearly perfect mode of the high finesse FPC.

Since the collision occurs at a slight angle
(see the mechanical description in section~\ref{section_meca} and in particular figure~\ref{cavity_geo}), the laser pulse duration must be close to that of the electron bunches, that is $\simeq 20$ps in our case at ATF. 


\begin{figure}[htbp]
\begin{center}
\includegraphics[width=0.8\textwidth]{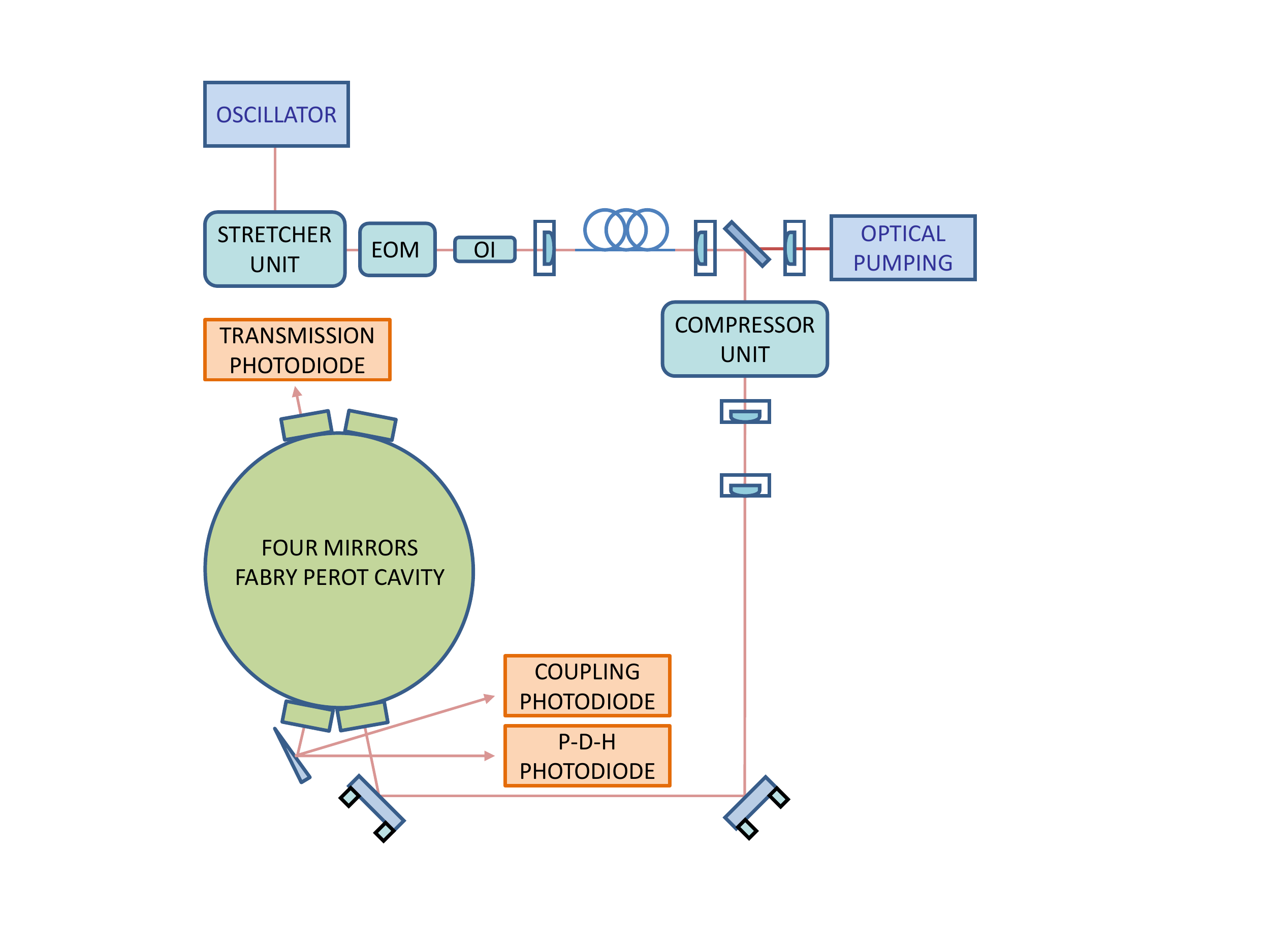}
\caption{Optical system including a frequency comb
  oscillator, a fiber amplifier and a FPC.}
\label{fig:schema_optique}
\end{center}
\end{figure}

\subsection{High average power low phase noise laser system}\label{section-opto}

The laser architecture has been dictated by the optical parameters
required for this experiment.

\begin{figure}[htbp]
\begin{center}
\includegraphics[width=0.8\textwidth]{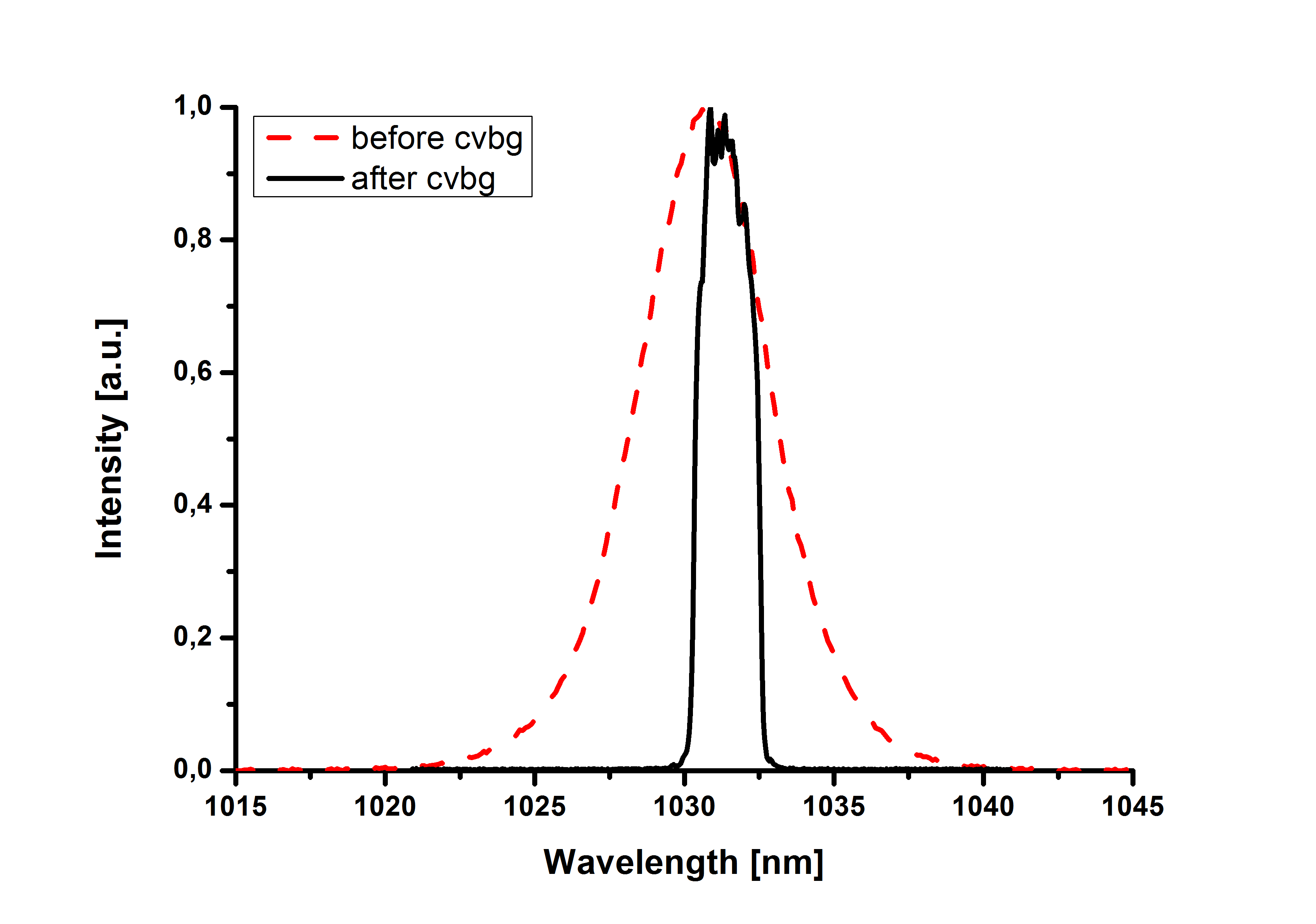}
\caption{Spectra of the laser beam measured before and after the CVBG}
\label{fig:transmissionCVBG}
\end{center}
\end{figure}

The laser system is based on a master oscillator power amplifier
(MOPA) implementing a chirp pulse amplification (CPA) scheme in
Yb-doped diode pumped fiber. It consist of a customized commercial
oscillator (Origami-10 from Onefive GmbH) delivering transform-limited
200 fs sech pulses at a wavelength around 1030 nm and at a repetition
rate of 178.5 MHz (to match half the bucket of the
DR). Because we make use of a fiber amplifier, amplification of
short pulses in a reduced diameter long waveguide may induce nonlinear
effects as the intensity in the fiber glass grows. The present
experimental setup and more specifically the external cavity
enhancement of the laser pulses simply can't suffer from any
non-linearities. In fact accumulating non-linear effect during the
beam propagation may result in amplitude-to-phase coupling potentially
ruining the locking procedure to the FPC. Therefore, to reduce
intensity in the gain medium, the pulses are temporally stretched
before amplification and recompressed after in the so-called CPA
architecture. In order to keep the system as compact as possible,
temporal stretching is performed by propagating the beam in a chirped
volume Bragg grating (CVBG~\cite{CVBG}) supplied by Optigrate. The principle of CVBG relies on the fact
different wavelenghts are reflected at different depth in the bulk
grating inducing a time delay between consecutive wavelengths of the
spectrum. By properly engineering the Bragg structure recorded in the
bulk material, it is possible to define a well controlled dispersion
law. In our case the dispersion is mainly of second order and
stretches the initial 200~fs pulses to more than 200~ps. The CVBG also
act as a filter as its transmission bandwidth is much more narrow than
than the oscillator bandwidth. This is depicted on figure
\ref{fig:transmissionCVBG} where we superimpose the oscillator
spectrum (5~nm) with the reflected spectrum (2.2~nm). The pulse
length required for the Compton scattering in our configuration are of
the order of several tens of ps and therefore perfectly compatible
with spectral filtering. Note here that filtering should not be too
important to allow sufficient stretching. The global efficiency of the
stretcher unit is about 37.5~\% mainly due to the filtering process~\cite{CVBG}.

\begin{figure}[htbp]
\begin{center}
\includegraphics[width=0.8\textwidth]{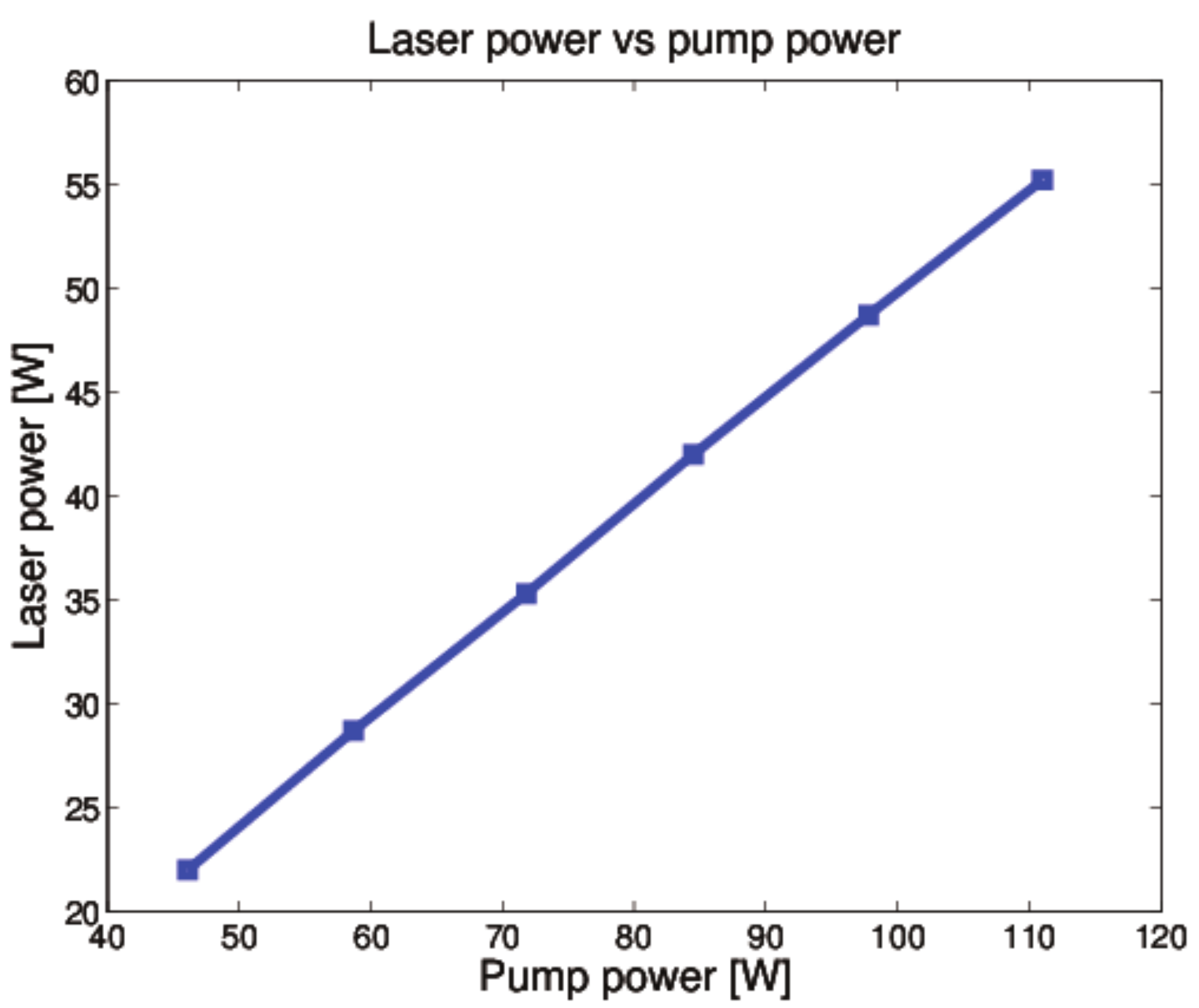}
\caption{Efficiency curve of the Yb-doped fiber amplifier.}
\label{fig:efficiency}
\end{center}
\end{figure}

\begin{figure}[htbp]
\begin{center}
\includegraphics[width=0.9\textwidth]{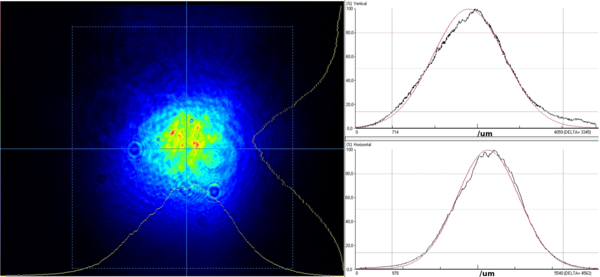}
\caption{Spatial intensity profile at the output of the fiber
  amplifier at 28 W output power}
\label{fig:spatial_mode}
\end{center}
\end{figure}

The stretched beam is then propagated through an electro-optic
modulator set for phase modulation and driven by a 5 MHz RF signal
necessary to generate the error signal for the PDH locking system described in section~\ref{section_electronique}.  In
order to prevent any artificial noise generation and locking
degradation from backscattered light, the laser oscillator is isolated
from the rest of the system and in particular from the amplifier with
a double Faraday isolator having an extinction ratio of 56 dB.  At
this point, the beam is injected in the fiber amplifier. The amplifier
consists of a 2-meter long double clad Yb-doped photonic crystal fibre
with a 40 $\mu$m core diameter and a 200 $\mu$m cladding diameter (NKT
DC-200/40-PZ-Yb supplied by NKT Photonics). The fiber is end-pumped by a high-power laser diode
(supplied by DILAS). The output power varies linearly with the pump power with a
slope efficiency of 51\% as shown on figure~\ref{fig:efficiency}.
A maximum output power of 55 W was achieved with this setup. The beam
quality is of major importance in the present context as any deviation
from a perfect Gaussian beam will be rejected by the high finesse FPC
and will not contribute to the power accumulation inside the
FPC. This is one of the reason why we have chosen amplification in a
single mode active fiber. 
We measured an amplified beam quality $M^2=1.07$ (axe x)	 $M^2=1.07$ (axe y)  in our laboratory. In figure~\ref{fig:spatial_mode} we give a
typical intensity profile measured at 28 W output power after installation of our system at ATF.
The beam is then sent in the compressor unit consisting of another
CVBG. In order to adjust the final duration, the compressor CVBG might
be different from the stretcher CVBG (still opposite sign but
different value). 
 In our case the minimal
duration is not an objective. Instead, we are more interested in
optimizing the duration of the laser pulse to match that of the electron bunch  and
maximize the gamma-ray production. In the present case, pulses are
recompressed to 68~ps and therefore exhibit a large chirp due to the
remaining second order dispertion. The chirp has no effect on the
gamma production nor on their characteristics. 

Finally, the beam is propagated through an arrangement of lenses
carrefully chosen to optimally match the FPC mode at injection. 

\subsection{High finesse Fabry-Perrot cavity}

\begin{figure}[htbp]
\includegraphics[width=.8\textwidth]{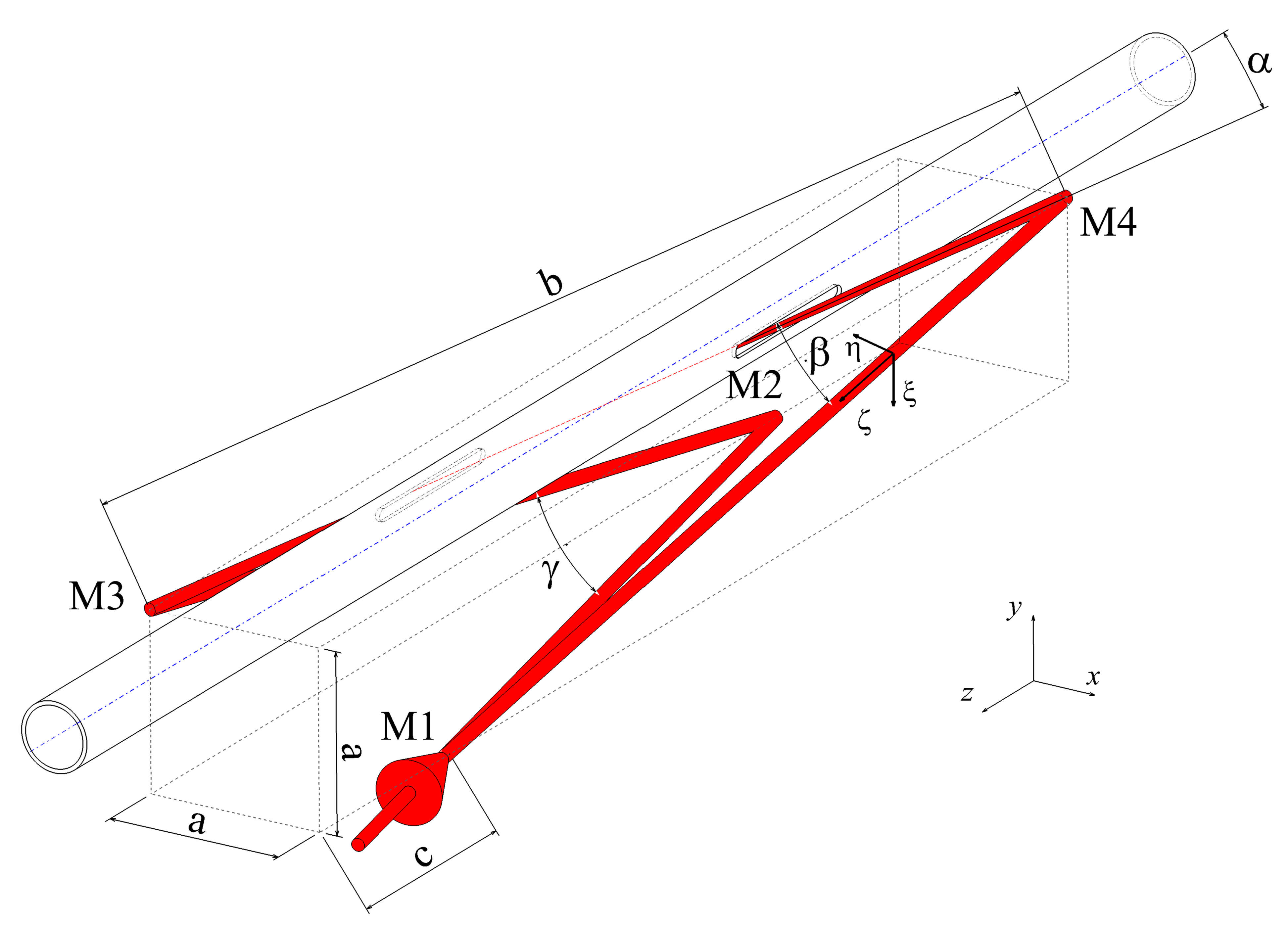}
\caption{Schematic drawing of the four-mirror cavity installed at ATF. The laser beam is represented by the red lines and the incident laser beam by a red arrow.  A rectangle parallelepiped, defined by $a=70$mm and $b=500$mm is drawn as a guide line. The spherical mirrors M$_3$ and  M$_4$ are located in the vicinity of  two of the corners of the parallelepiped. The two flat mirrors are shifted by $c \approx 81.19$mm along the bottom line segments  M$_1$ and  M$_2$. The beam pipe and the rectangular slit of 5mm large is shown as well as the fixed reference frame $(x,y,z)$ and the frame attached to the laser beam optical path $(\xi,\eta,\zeta)$. 
$\alpha=8.05^o$ is the laser beam/electron beam crossing angle and $\beta/2=6.25^o$, $\gamma/2=7.6^o$ are the two incident
angles on  M$_2$ (and M$_1$) and  M$_4$ (and M$_3$) respectively.  }
\label{cavity_geo}
\end{figure}

As mentioned in the introduction, two-mirror cavities are unstable for
small mode waist sizes. To address this issue and others we choose to
use a four-mirror cavity instead. In order to provide stable
circularly polarized eigenmodes (as required by high energy physics~\cite{araki,goodie}), a non-planar tetradedron geometry is chosen
\cite{nous4mirroir} as shown in figure \ref{cavity_geo}.  The cavity
is made of two concave mirrors, M$_3$ and M$_4$ of radius of curvature
0.5m and of two flat mirrors M$_1$ and M$_2$. The cavity finesse is
determined by the reflection coefficients $r_i$ of the mirrors
$M_i$ which depend themselves on the number of SiO$_2$/Ta$_2$O$_5$
double layers deposited on the mirror substrate. In order to optimize
the laser/cavity mode coupling, we fixed $r_1\approx r_2r_3r_4$
\cite{hello}, where M$_1$ is the entrance cavity mirror. For the
cavity commissioning we choose
$|r_2|^2\approx|r_3|^2\approx\|r_4|^2\approx 1-330$ppm and
$|r_1|^2\approx 1-1060$ppm leading to a finesse
$F=\pi/(1-|r_1r_2r_3r_4|)\approx 3000$ ({\it i.e.} a power gain of
$G=F/\pi\approx 1000$). The mirrors coating were made by ourselves at
the LMA Laboratory of Lyon.  In the finesse expression given above,
the losses \cite{hello} are neglected, this is justified because we
have measured the absorption and the scattering of the mirror coating
to be $\approx 0.6$ppm and $\approx 3$ppm respectively.

\begin{figure}[htbp]
\includegraphics[width=1.\textwidth]{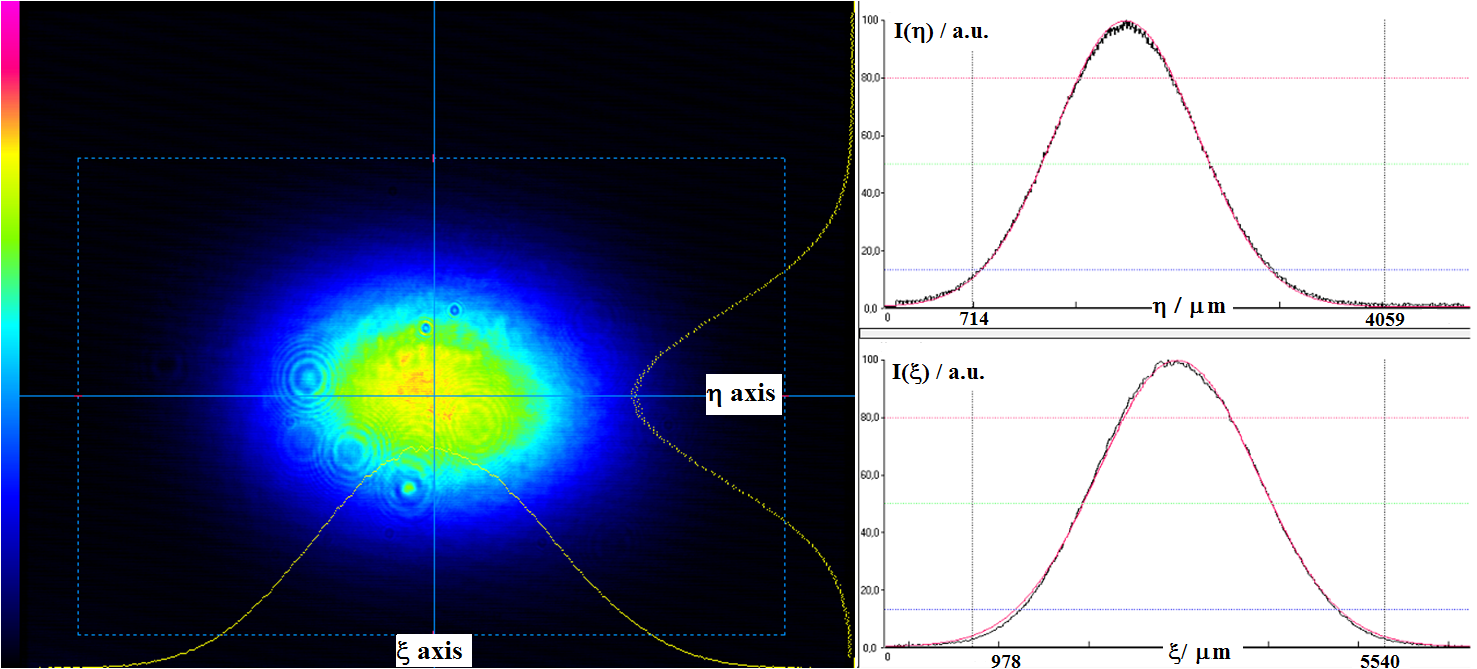}
\caption{Typical beam intensity transverse profile $I(\xi,\eta)$ as measured by a laser beam scanner located just behind M$_2$. The left plot shows the two dimensional as a function of the transverse coordinates $\xi$ and $\eta$. The right plots show the projections along the principal axes $\xi$ and $\eta$ together with Gaussian fits. From these fits we obtain $\omega_\xi=1.75$mm and $\omega_\eta=1.17$mm.}
\label{beam_profile}
\end{figure}

The mirror positions (see figure  \ref{cavity_geo}) are chosen such that the cavity round trip optical path matches the distance between two electron bunches of the ATF, that is $\approx c/ 178.5MHz\approx 1.681$\ m. The FPC being non-planar, the eigenmodes belong to the class of general astigmatic beam \cite{kogel_arnaud,arnaud}, that is the intensity spatial profile is an ellipse with principle axes rotating during the propagation (see \cite{papier_Al} for a tetahedron cavity). The distance between the spherical mirrors is tunned in order to provide a cavity mode waist sizes of $52\mu$m and  $76\mu$m ({\it i.e.} equivalent  Gaussian  intensity spot radii of  $26\mu$m and  $38\mu$m) half way between the two spherical mirrors. With such beam radii, diffraction losses induced by the beam pipe slit aperture (see section \ref{section_meca}) are negligible.
Figure \ref{beam_profile} shows the spatial beam profile measured behind M$_2$. From this measurement, using the eigenmode calculation of \cite{arnaud} we can estimate the evolution of the beam size inside the cavity. 
The effect of the ellipse axis rotation was studied in \cite{papier_Al}, in our setup it also leads to negligible luminosity loss. 
We checked our calculations in a dedicated experiment where the cavity beam waist was imaged using an converging lens located in transmission on M$_4$\cite{these_yasmina}. A good agreement was obtained. The circular polarization of the eigenmodes were measured in \cite{honda}.
We also checked numerically that, for the ATF electron beam parameters~\cite{ATF}, the laser beam intensity profile rotation has a very small impact of the electron/laser beams luminosity~\cite{papier_Al}. We verified numerically and experimentally that varying the distance between the two flat mirrors within $\pm 2$mm, that is the dynamical range of the mirror longitudinal positioning system, the beam waist size does not change significantly.  On the other hand, the  beam waist size is very sensitive to the distance between the two spherical mirrors.   

\section{Mecanical design and construction}  \label{section_meca}
\subsection{Requirements for the mechanical design}
The strongest requirements on the design of a non planar cavity (see section \ref{section_opto}) for Compton scattering come from its installation in an accelerator. As the ATF is one of the electron rings with the world's smallest transverse emittance~\cite{LowEmittanceATF} we had to minimize any perturbation on the electron beam by providing a sufficiently low vacuum (below $10^{-7}$mbar) and by ensuring as little discontinuity  as possible of the beam pipe impedance. The later point means that the beam pipe must continue inside the cavity.
To allow the laser beam to cross the electron beam a slit had to be machined in the beam pipe. This slit has a vertical dimension of 5mm (see figure \ref{cavity_geo}), the largest acceptable  aperture from the ATF machine point of view. Such a small inner aperture requires the ability to tilt the cavity mirrors so that the laser beam can be steered through this aperture while being sufficiently immune to environmental noises. This must be achieved by using sub-micrometric actuators fixed on gimbal mirror mounts.

Since the laser beam is pulsed, the cavity round trip must coincide with the distance between two electron bunches at the ATF. Some of the mirrors must thus also be mounted on longitudinal translation stages for a coarse tuning of the optical path and one of them must be connected on a piezo electric transducer to allow  fine tuning ({\it i.e.} for the synchronization with the ATF clock, see section \ref{section_electronique}). 


The 5mm beam pipe aperture slit fixes the divergence of the optical cavity mode, {\it i.e.} the  laser beam waist size avoiding diffraction losses is limited and its minimum value is estimated to $\approx 35\mu$m. In order to optimize the luminosity it is then necessary to reduce the laser beam - electron beam crossing angle. The mirror mounts of the two spherical mirrors must then be designed in such a way to minimize the material budget between the mirror center and the beam pipe. 

Finally, one needs to provide a high level of mechanical stability for the synchronization of the laser oscillator with the high finesse cavity. Special attention must thus be paid to the stability of the mirror mounting system and to the isolation of the whole optical system against thermal fluctuations and, ground and air induced vibrations.

 \subsection{Description of the mechanical system}

\begin{figure}[htbp]
\includegraphics[width=.6\textwidth]{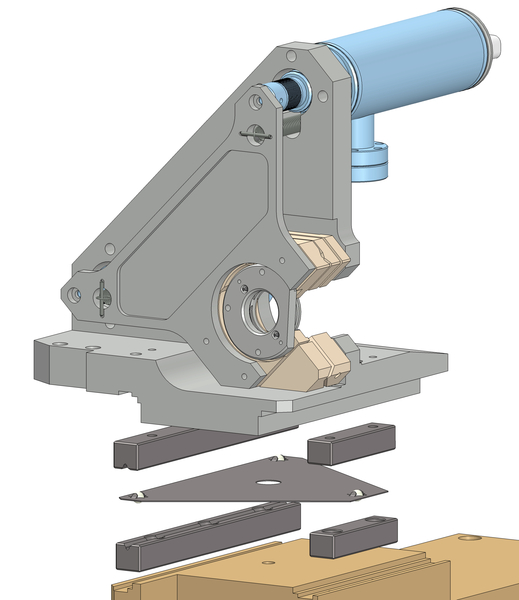}
\caption{Technical drawing of a spherical mirror mounting system. The longitudinal translation stage, made of three ceramic balls, two guiding rails and a triangular plate is shown as an exploded view.}
\label{global_mount}
\end{figure}

\begin{figure}[htbp]
\includegraphics[width=1.\textwidth]{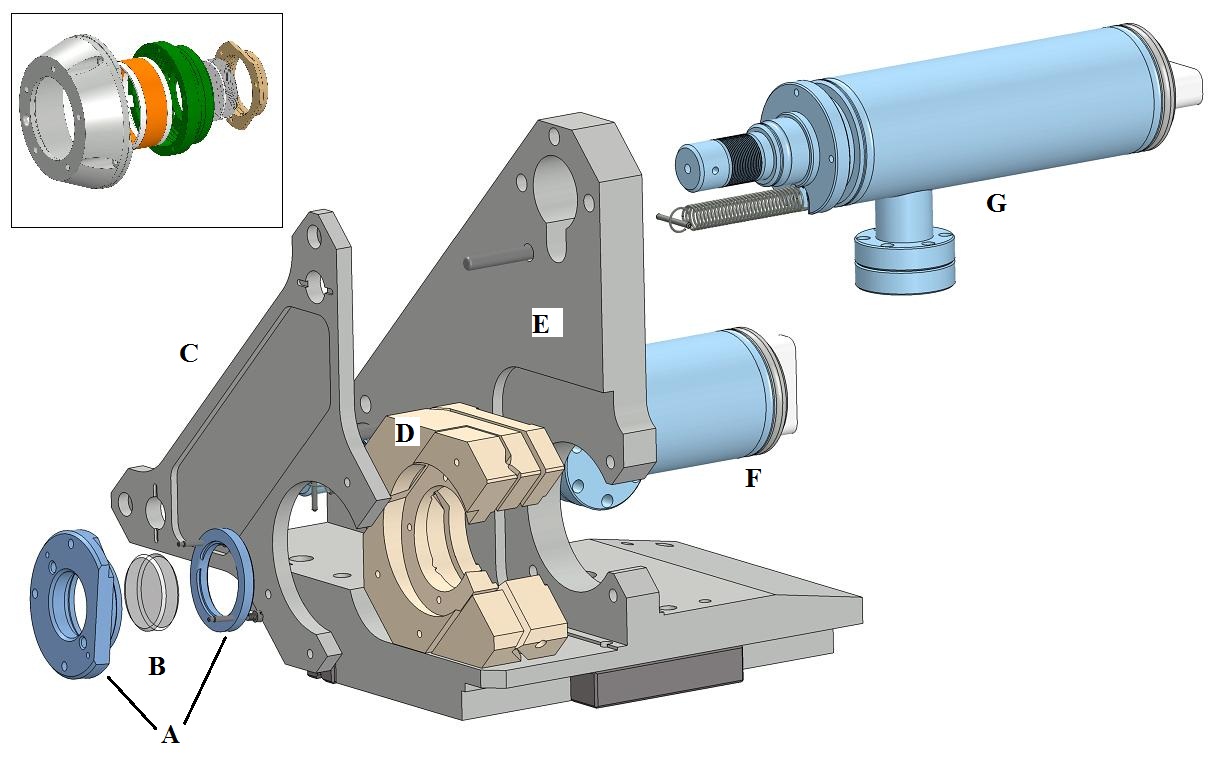}
\caption{Exploded view of the tilt system. A: mirror ring system; B: cavity mirror; C: triangular arm; D: gimbal mirror mount; E: mirror mount support; F: encapsulated motor for actuating the rotation along the $x$ horizontal axis; G: encapsulated motor for actuating the rotation along the $y$ vertical axis. On the top left of the figure, the mirror ring containing the annular piezoelectric transducer (represented by an orange cylinder) is also shown. }
\label{tilt_system}
\end{figure}

\begin{figure}[htbp]
\includegraphics[width=.8\textwidth]{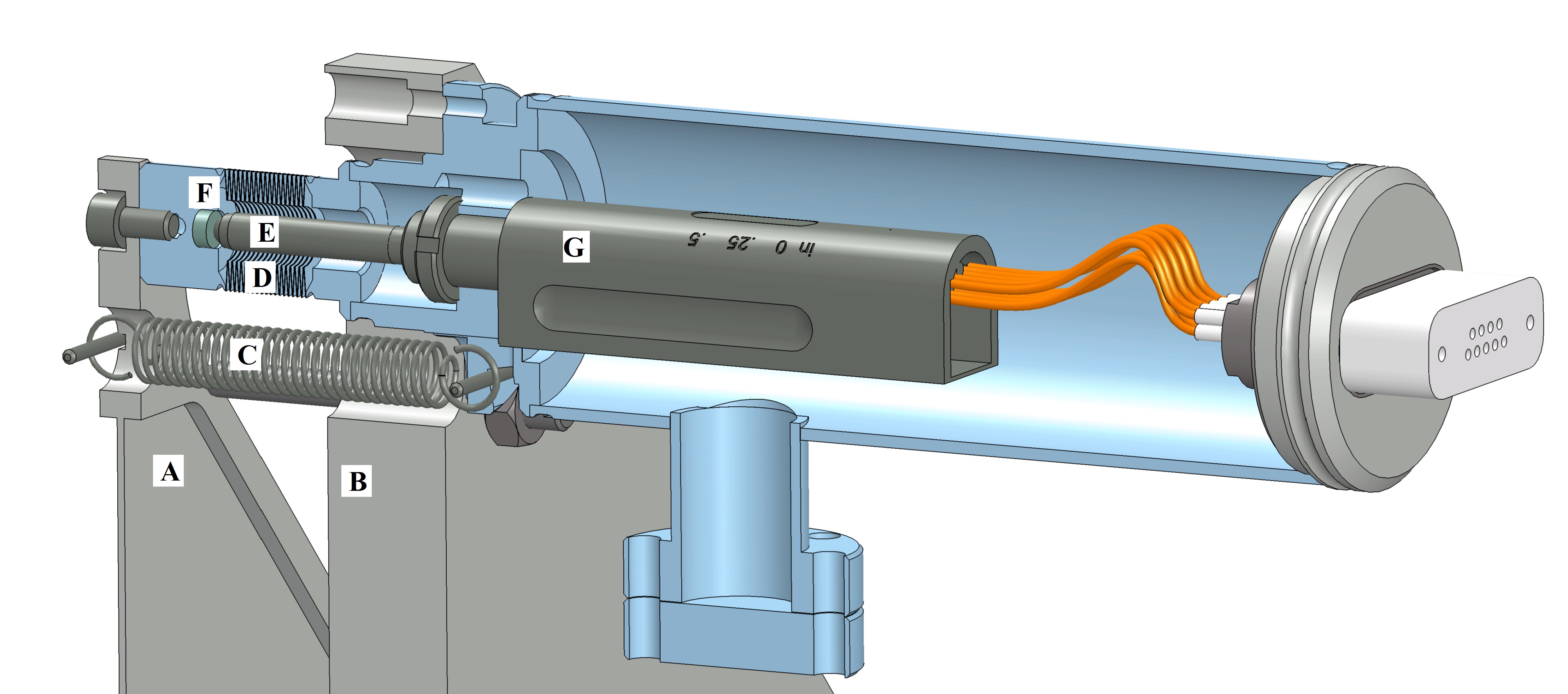}
\caption{ Cut view of the encapsulated commercial stepper motor. A: triangular arm; B: mirror mount base; C: spring; D: small bellow; E: stepper motor screw; F: hard steel pellet used to avoid friction of the end of the motor screw; G: stepper motor body.}
\label{encapsulation}
\end{figure}

Each mirror is mounted as described on figure \ref{global_mount}. Three actuators allow to tilt the mirrors in $x$ and $y$ directions and to move them along the $z$ axis.  An exploded view of the tilting system is shown on figure \ref{tilt_system}. The mirrors are inserted in a ring fixed on a gimbal mount made of two flexible hinges. The hinges are machined by wire erosion-cutting of a stainless steel cube and the hinge motion results from the elasticity of the remaining thin material: one of the hinges can be tilted in the $x$ direction and the other in the $y$ direction while keeping the mirror center at the same location.  
To control the hinge tilts, two encapsulated stepper motors (see also the exploded view on figure \ref{encapsulation}) are connected to a triangular arm, itself fixed to the surface of the gimbal mount. Then a spring is used to constrain the system. Figure \ref{encapsulation} shows that the motor motion is transmitted to the triangular arm through a small bellow  and that a hard steel pellet is fixed at the end of the cap in order to reduce friction effects. The motor cap is filled with Helium through a small flange in order to identify possible leaks. For the minimum motor step $\pm 0.3\mu$m we have measured a tilt of $\pm 2.7\mu$rad of the mirror plane. The maximum motor excursion of $\pm 1$mm therefore corresponds to a maximum tilt of $\pm 9$mrad.  Figure \ref{global_mount} shows in fact the mounting system of one of the two spherical mirrors. The side cut of the mirror mount, leading to a reduced laser-electron beams crossing-angle, is clearly visible.      
 For one of the flat mirrors, the ring was adapted to include an annular piezo electric transducer (see figure \ref{tilt_system}) constrained by a flexible ring made by wire erosion-cutting. 

\begin{figure}[htbp]
\includegraphics[width=1\textwidth]{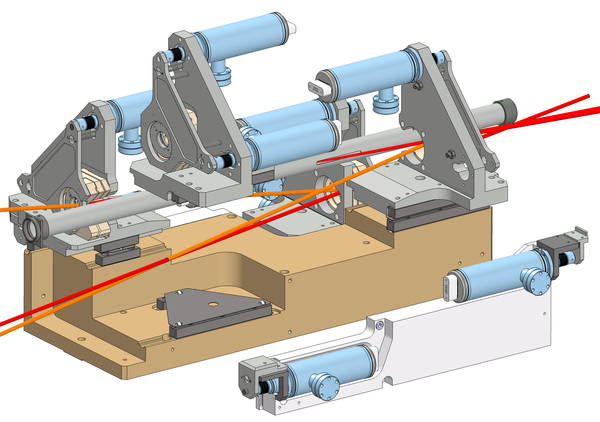}
\caption{Exploded view of the four mirror mounts system. The invar plate with two  encapsulated motors and one of the flat mirror mount system have been unlinked from the heavy stainless steel base for the drawing.  }
\label{mechanic_global}
\end{figure}

\begin{figure}[htbp]
\includegraphics[width=.8\textwidth]{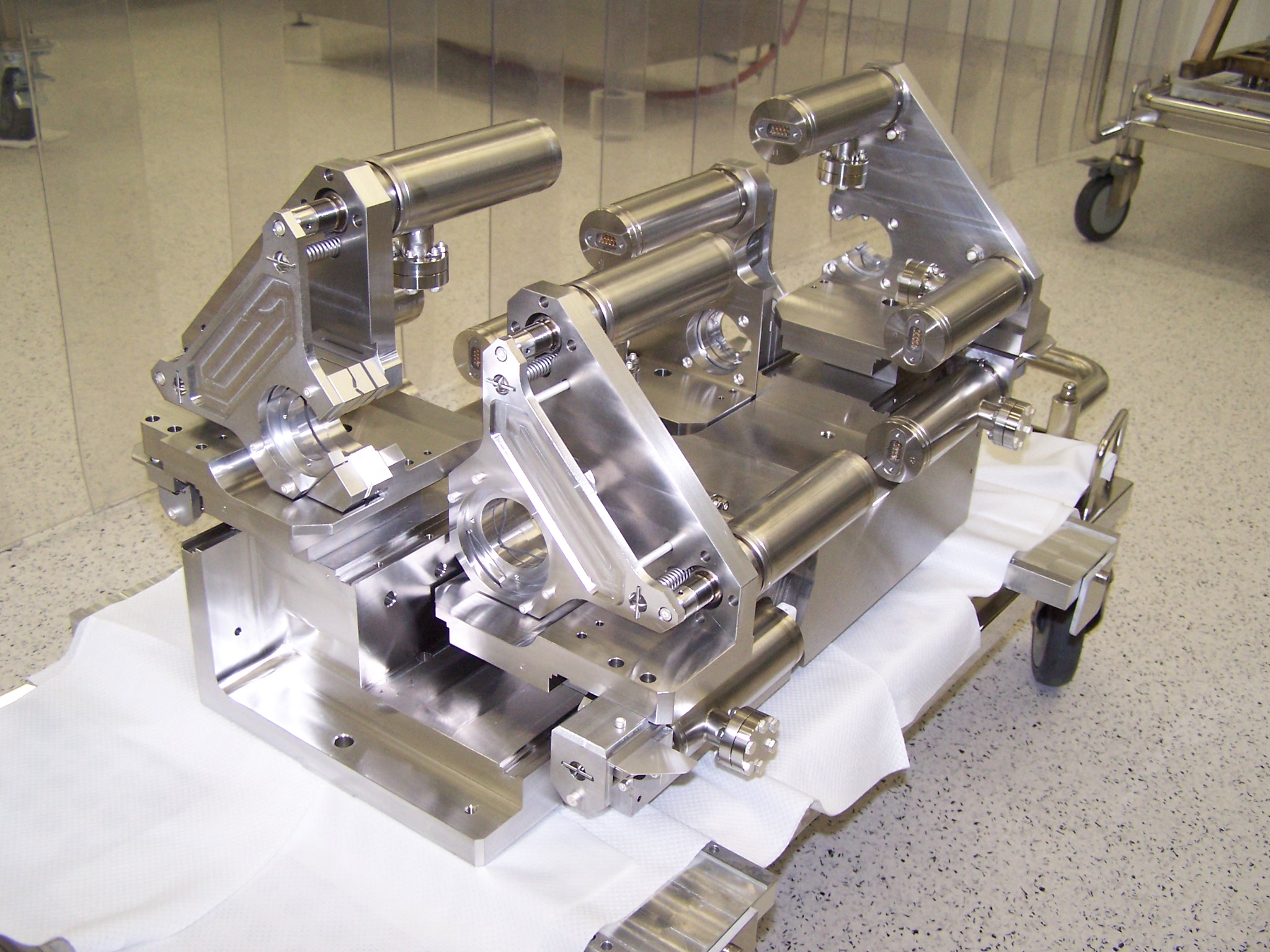}
\caption{Picture of the four-mirror mounting system taken during the assembly in a clean class 10 (ISO4)  room.}
\label{photo_mirror_mount}
\end{figure}

The longitudinal translation stage is shown on figure \ref{global_mount}. The mirror mounting system is put on three ceramic balls guided by two tracks. The motion is also done thanks to an encapsulated motor. As shown on figure \ref{mechanic_global}, the motors are fixed on an invar plate itself fixed by its center on a 65~kg stainless steel base plate in order to increase the stability of the mirror positions under thermal variations. A picture of the four-mirror system is shown in figure \ref{photo_mirror_mount}.
 
Finally the whole system is put inside a cylindrical vacuum chamber 
 supplied by Caburn (MDC). Because of the complexity of the system no baking was done. Therefore all pieces were carefully cleaned and mounted inside the chamber in a class 10 clean room ({\it i.e.} ISO4) at LAL, Orsay, then sealed and sent to KEK. 
Once at the ATF, all manipulations inside the cavity were done under a class 100 ({\it i.e.} ISO5) clean laminar air flow. We measured the position of each element inside the chamber with an absolute precision of 0.2mm so that we were able to fix $z$ position variation range of each mirror mounts to $\pm1$mm.
 

Because of the high vacuum requirements, all mechanical elements are made of stainless steel (and all screws and nuts are silver coated). 
The special gimbal mirror mount and the three ceramic balls system for the longitudinal motion are designed to avoid galling, making the mirror motion frictionless.

 By using encapsulated motors inside the vacuum vessel we were able to decouple the tilt from the translation mirror motion and to reduce the effects of external noises. The equilibrium vacuum measured inside the cavity is about $3 \times 10^{-8}$mbar ($\approx$30m of high vacuum electrical wires are contained in the vacuum chamber) and no noticeable contamination from atoms with Z>18 has been measured as specified by the ATF vacuum requirements and to preserve the high reflectivity of the cavity mirror coatings. 

A set of isolators are used to reduce the effects of environmental noises: two bellows are located on each side of the vacuum chamber to align and isolate the system from the rest of the damping ring; three A5 aluminum feet inserted between the chamber and the optical table; and the optical table is standing on a dozen of elastomer bumpers. The optical table is enclosed in a box made of an acoustic insulating material and a system of heating wires is used to stabilize the temperature inside this box with a precision of $\pm0.05^o$C. 

\clearpage
\section{Fabry-Perot Cavity stabilization and the electronic system}  \label{section_electronique}

\begin{figure}[htbp]
\begin{center}
\includegraphics[width=.6\textwidth]{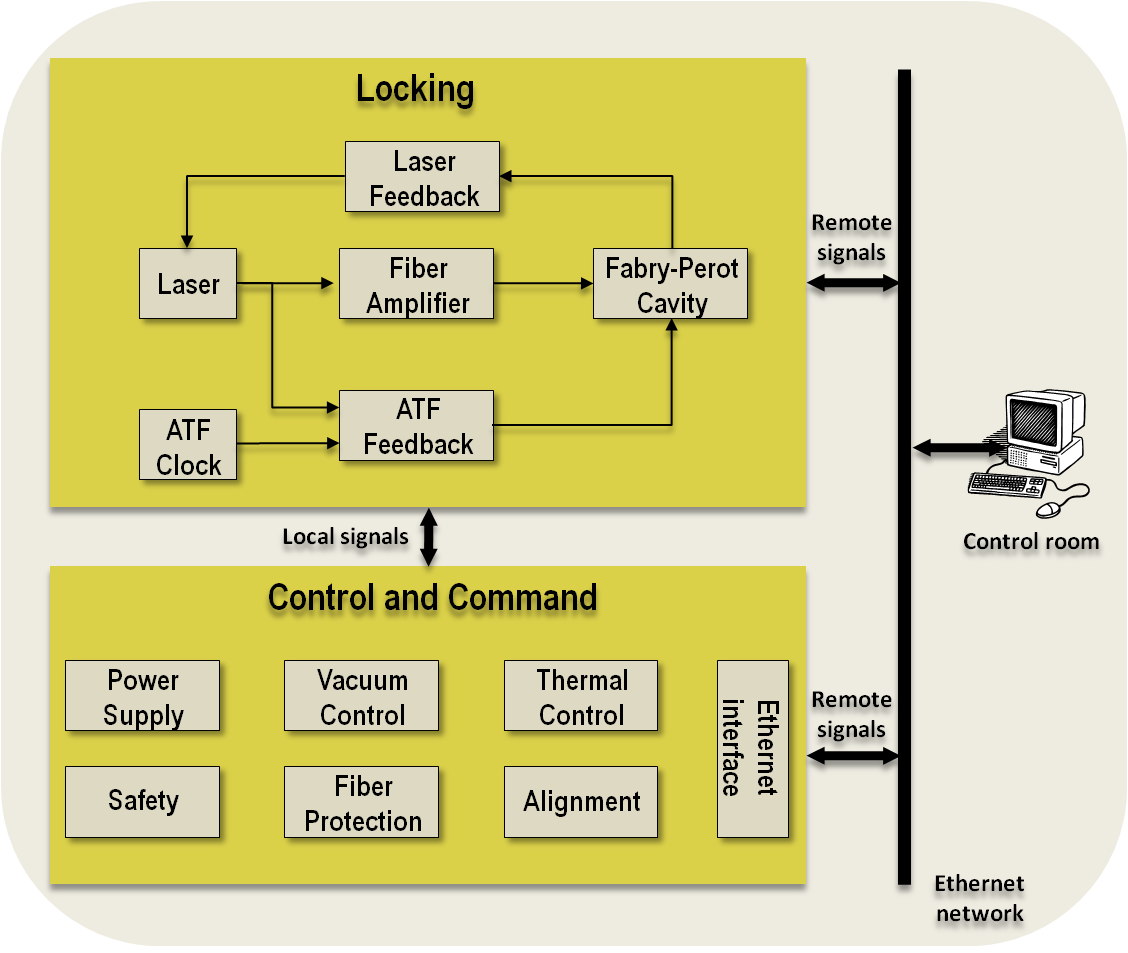}
\caption{Overall architecture of the active stabilization system. The upper part is performed by the FPGA and the lower part is done by several remoetly controlled sub-systems.}
\label{elec_architecture}
\end{center}
\end{figure}

Although the utmost care has been taken with the design, construction and assembly of the FPC this is not sufficient to reach the high level of stability required for a high gain in the cavity and to achieve collisions with the electrons circulating in the ATF damping ring. Therefore we have equipped our FPC with an active stabilization system. This active stabilization system performs several tasks. First of all it must ensure that the duration of a round trip for the photons circulating in the FPC has an integer relation with the time between two electrons bunches. The second task of the active stabilization system is to ensure that the laser pulses injected in the FPC arrive in time. 
The third task of this system is to  have a slow control  on the environmental variations. The overall architecture of the system is shown on figure~\ref{elec_architecture}.


\begin{figure}[htbp]
\begin{center}
\includegraphics[width=.6\textwidth]{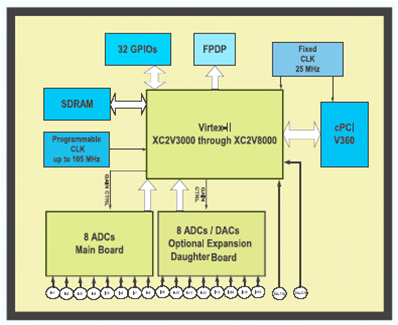} 
\caption{Lyrtech board architecture}
\label{elec_lyrtech}
\end{center}
\end{figure}

\begin{figure}[htbp]
\begin{center}
\hspace*{-2.5cm}\includegraphics[width=1.\textwidth]{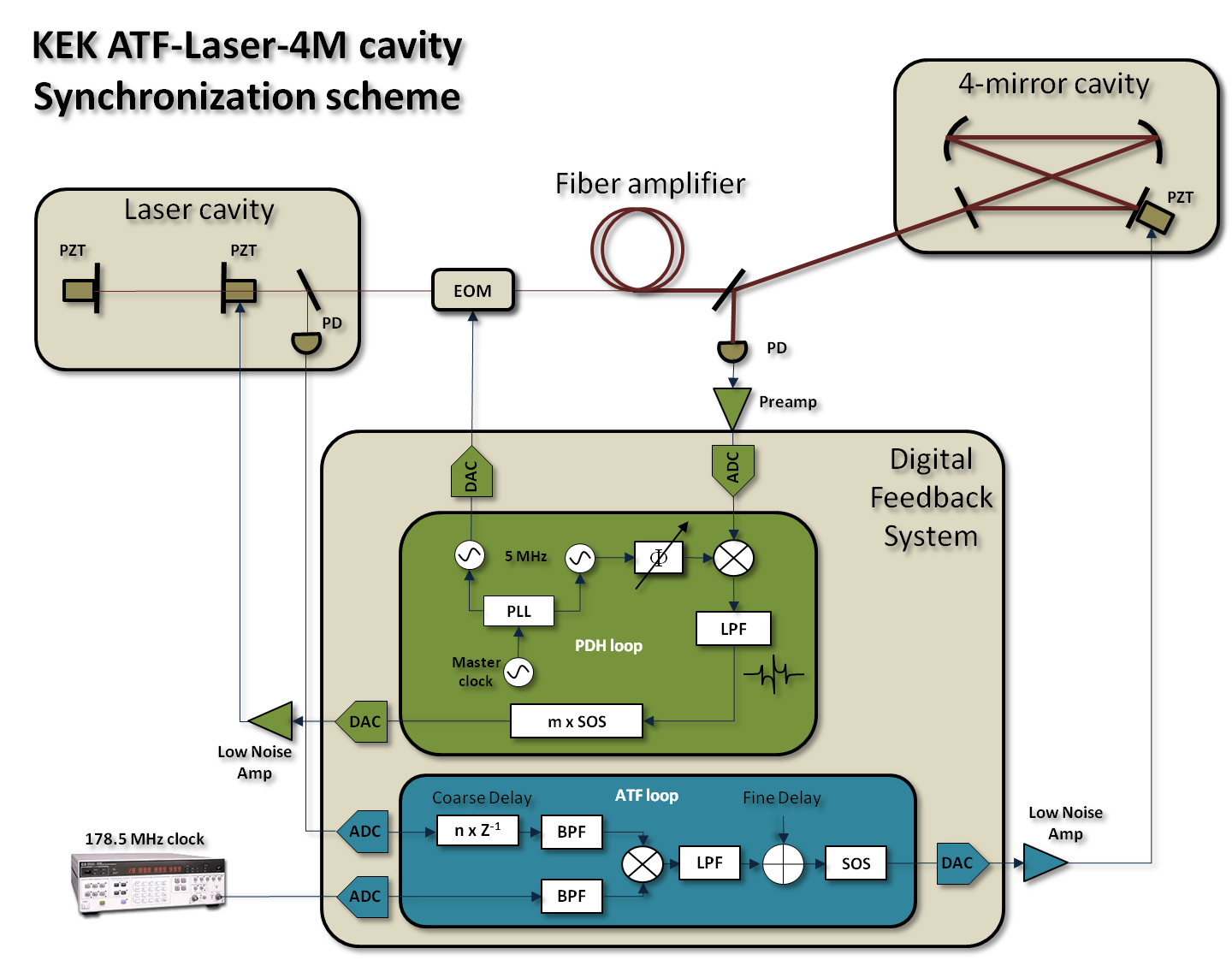}
\caption{Implementation of the digital feedback system. The blue area is a digital phase lock-loop used to synchronize the FPC on the ATF clock. The signal produced by the feedback loop is used to adjust the length of the FPC by changing the tension applied on a PZT behind mirror M1. The green area is an implementation of the PDH technique to synchronize the laser oscillator on the FPC. When the photodiode located behind mirror M2 receives enough transmitted power the error signal produced using the PDH technique is used to vary the length of the laser oscillator cavity by acting on a PZT. All these features are implemented in a single FPGA board.}
\label{fig:elec_full_system}
\end{center}
\end{figure}

The first two tasks of the active stabilization system have to be performed at a very high speed to have a fast response time.
 We have therefore chosen to implement it on a VIRTEX-II FPGA (XILINX) commercial board VHS-V2-ADAC from Lyrtech.  The overall board is shown on figure~\ref{elec_lyrtech}.
 This board is using several 14 bits ADC and DAC with a sampling speed of 60 megasample/s and clocked at 60~MHz. 
This board implements a double feedback system that synchronizes the FPC with the ATF clock and the laser oscillator with the FPC. 
The implementation of this double feedback system is shown on figure~\ref{fig:elec_full_system} and is detailed below.
The other task of the active stabilization system, the correction of slow environmental variations, is performed by several sub-systems controlled remotely over the network by a desktop computer.

\begin{figure}
\begin{center}
\includegraphics[width=0.8\textwidth]{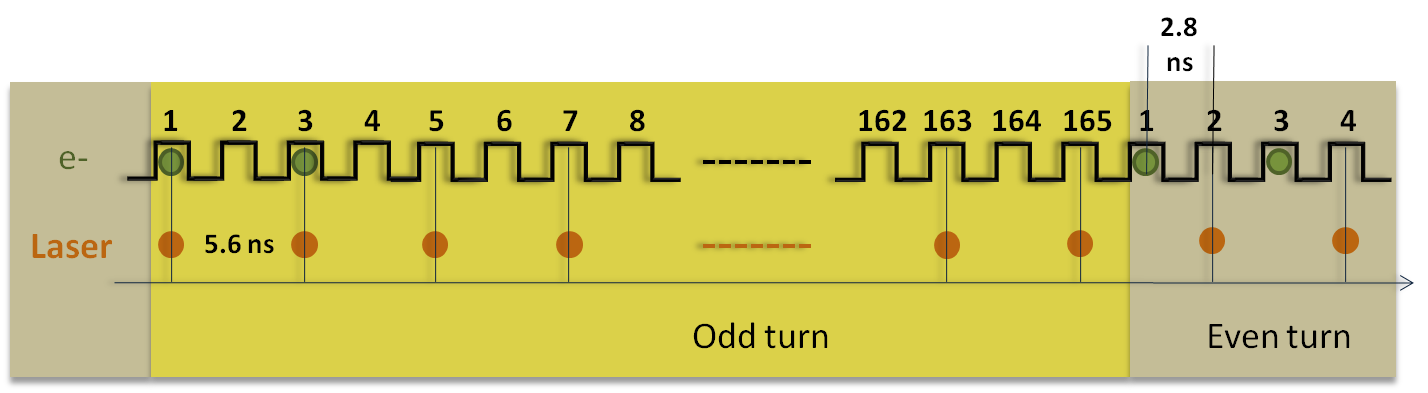}
\caption{Time structure of the RF buckets (electron bunches) and of the laser, showing that collisions occur only every other turn.}
\label{elec_bunch_structure}
\end{center}
\end{figure}

 Ideally one would like the round trip of the photons in the FPC to be exactly equal to that of the electrons in the damping ring. However this would have required to match the length of the FPC to that of the damping ring, 154~meters, which would have been unpractical.  Furthermore the ATF can operate in multibunch mode and in that mode bunches can be separated by 2.8~ns or a multiple of that value. As a round trip of 2.8~ns would have been impractically short, we choose to give our cavity a round trip of 5.6~ns (that is 1680~mm or 178.5~MHz).
Given that the ATF has an harmonic number of 165 (i.e. it contains 165 RF buckets separated by 2.8~ns each), our photons perform 82.5~round trips in FPC while the electrons perform one round trip in the cavity. As a consequence a given electron bunch will collide with the photon pulse stored in the cavity on every other turn and during that time the photons will have done 165 round trips in the FPC. This is illustrated on figure~\ref{elec_bunch_structure}.

Our system has to be able to work
in the case where the two pulses (laser and electrons)  are as short as about 20 ps every 5.6~ns, therefore the relative stability has to be around $10^{-3}$, the natural stability of both systems allows us to use a feedback with a low bandwidth (about 100~Hz). Thus to lock the FPC round trip on the ATF clock we decided to act on the length of the FPC which has larger mirrors and thus are slower to react. For the other lock, the noise on the length difference  between the FPC and the laser cavity has to be better than the wavelength over the finesse (the relative stability has to be around $10^{-10}$), requiring a high feedback bandwidth and thus this loop acts on the mirrors of the laser cavity which are smaller.

\subsection{FPC synchronization with the accelerator}

A digital phase lock loop (PLL) is used to ensure that the round trip frequency in the FPC (178.5~MHz) keeps its integer relation with the round trip in the damping ring (2.165~MHz) with a very high accuracy (jitter of about 10~ps RMS). 
A synthesizer (AGILENT E8663B) locked to the ATF master clock is used to generate a 178.5~MHz reference signal. This reference signal is feed to an ADC where it is digitized with a sampling rate of 60~MHz. At this sampling rate the reference signal has a folded frequency of 1.5MHz ($178.5\mbox{MHz} - 3 \times 60\mbox{MHz}$).  
A photodiode located near the exit of the laser oscillator is used to generate a second signal (see figures~\ref{fig:schema_optique} and~\ref{fig:elec_full_system}). This signal is also digitized at the same rate.
These two digital signals pass through finite impulse response filters (FIR) acting as a band-pass filters. They are then mixed together. The resulting signal contains two components: one at high frequency (about 3~MHz) that is the sum of the two input frequencies and one at low frequency (near DC) that is the difference between these two frequencies. The high frequency component is removed by a digital low-pass filter and the remaining (low frequency) signal contains information on the frequency and phase difference between the two input signals. If the two signals were at the same frequency but at a different phase the resulting signal will be a simple DC offset whose value is proportional to the phase difference between the inputs.
This signal is used to slightly move longitudinally mirror M2 in the FPC by varying the tension applied to a piezo-electric transducer (PZT) and therefore change the round trip (by 4.5nm/V) and the frequency of the FPC. This PZT has a low dynamic range and is used with a bandwidth of about 100~Hz.
 Additional displacements can be achieved thanks to motors located behind each mirror of the FPC. These motors allow changes in length of the FPC of the order of 25~mm, that is $\pm 130 kHz$.
This digital PLL is illustrated on the green part of figure~\ref{fig:elec_full_system}.

Once the two signals are synchronized by the PLL their relative phase can be adjusted in a coarse manner by delaying the signal coming from the laser by a given number of FPGA clock cycles or in a fine manner by adding a given value to the output of the low-pass filter  after the mixer.

The fact that the sampling frequency of the FPGA is lower than the frequency of the signals to be digitized is not a problem as both signal are almost at the same frequency and therefore the existence of phase and frequency differences between the two analog signals will lead to phase and frequency differences in the digitized signals. When the two signals have the same phase and frequency their digitized images are also similar and therefore no correction is applied to the FPC. We can ignore the risk of having the two signals being harmonic of each other when sampled at 60~MHz because this would result in a significant change in the length of the FPC or the damping ring (at least 60~MHz $\times c$) which are not physically possible.

\subsection{Oscillator synchronization with the FPC}

To ensure that the laser pulses are properly coupled to the FPC they must be injected at the correct time. To ensure this we use the Pound-Drever-Hall (PDH) technique~\cite{PDH} to adjust the length of the laser oscillator. The optical frequency stabilization must be better than the repetition rate over the Finesse (the FPC linewidth), which is 60 kHz. This is equivalent to a stabilization of the 1680~mm   FPC length better than the laser wavelength over the finesse: 0.34~nm.

\begin{figure}[htbp]
\begin{center}
\includegraphics[width=.8\textwidth]{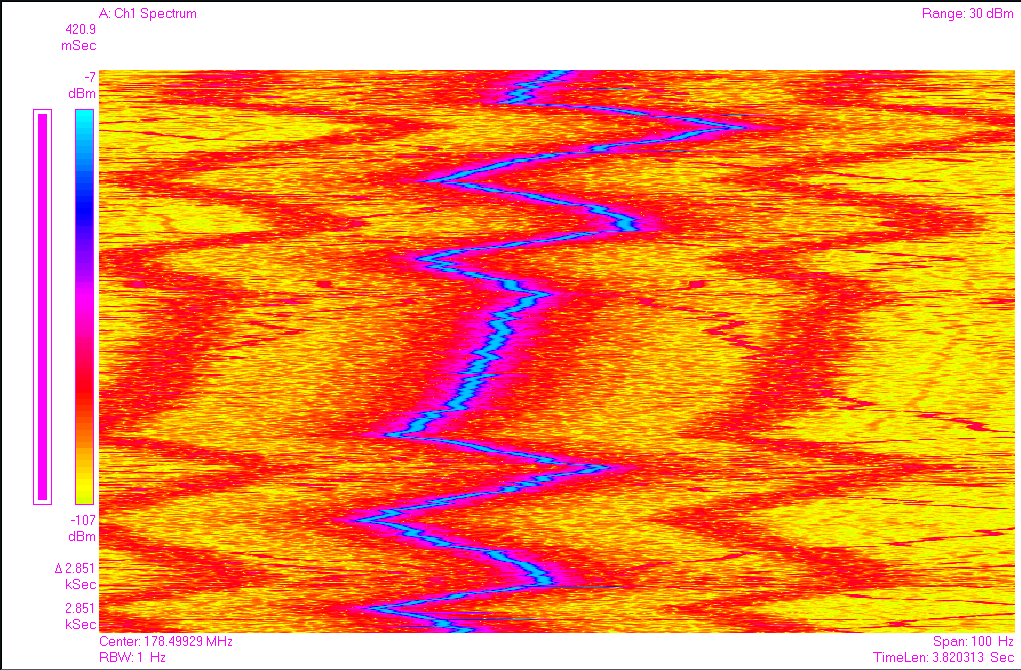} 
\caption{Long term stability drift of the OneFive Origami at 178.5MHz repetition rate.  The horizontal axis is frequency offset (horizontal SPAN = 100 Hz) and the vertical axis is the measurement duration (47~minutes).  The color of each pixel of this plot gives the power measured at the time and frequency indicated by the axis, red being the most intense. The red line on the plot shows the variation of the carrier frequency and the two light blue lines show that the sideband frequencies follow a similar pattern.}
\label{elec_long_osc_stability}
\end{center}
\end{figure}

\begin{figure}
\begin{center}
\includegraphics[width=1.\textwidth]{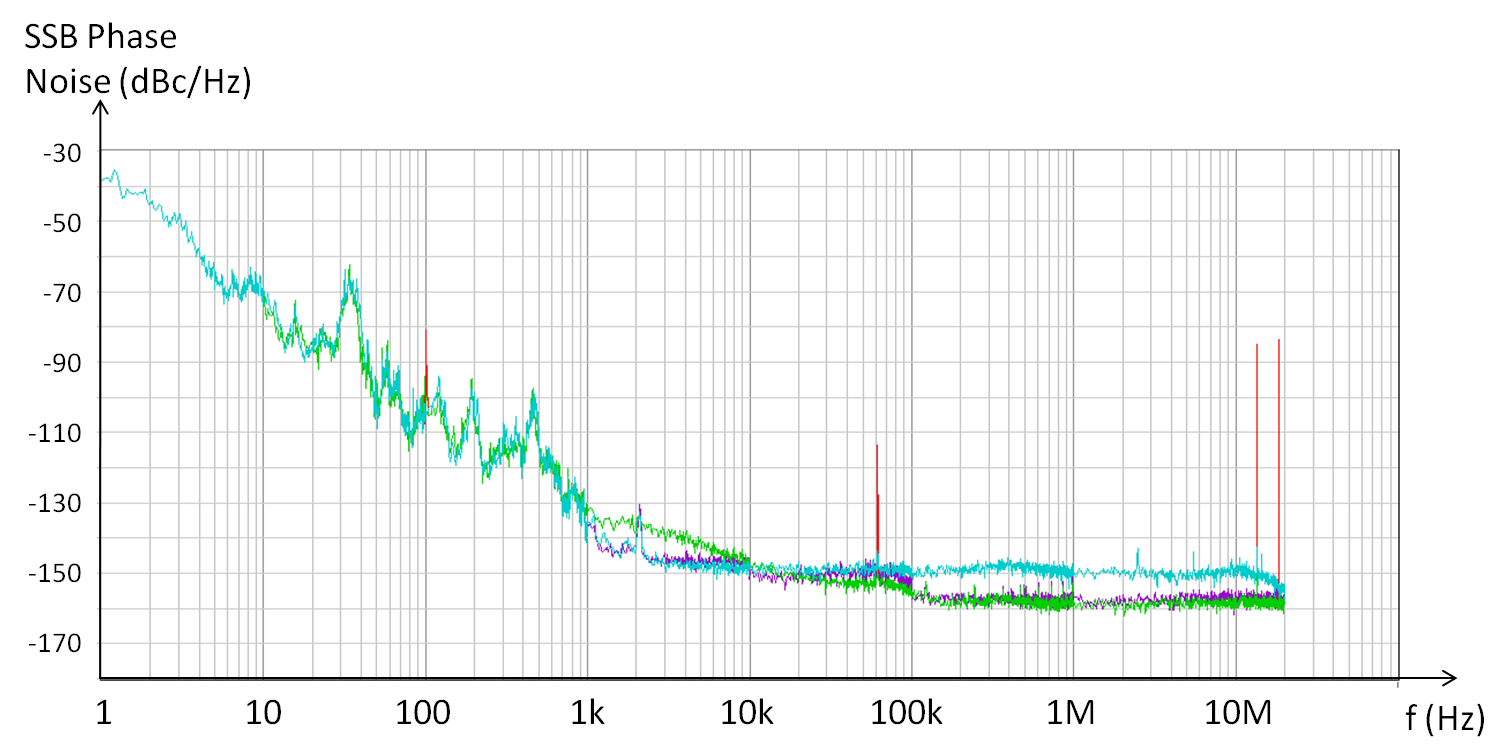} 
\caption{Phase noise power spectrum L(f) (Frep =  178.5 MHz) measured with different setup:  internal InGaAs photodiode and 20dB amplifier (blue),  external photodiode (Thorlabs DET10A/M; green) and external Hamamatsu InGaAs photodiode Hamamatsu and OneFive amplifier (red).}
\label{elec_osc_phase_noise}
\end{center}
\end{figure}

The natural stability of the laser oscillator is shown in figures~\ref{elec_long_osc_stability} and \ref{elec_osc_phase_noise}.

\begin{figure}
\begin{center}
\includegraphics[width=.8\textwidth]{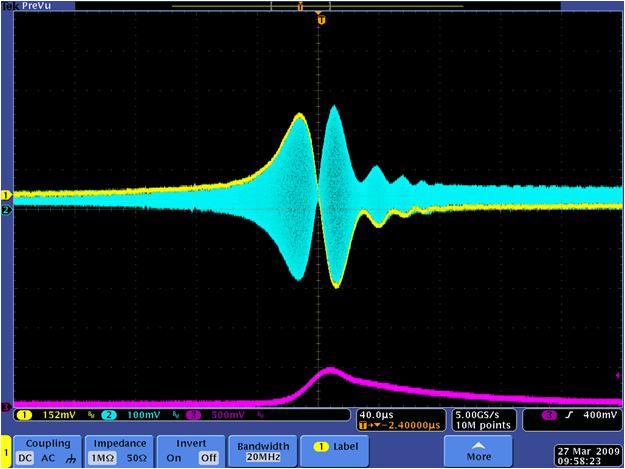} 
\caption{PDH error signal (blue) and IF signal (yellow) observed when the length of the FPC is varied with time. In the center one can clearly see the area where the error signal is linear. The purple line shows the power transmitted by the FPC.}
\label{fig:elec_DPDH_plot}
\end{center}
\end{figure}

As the PDH technique is described elsewhere in the literature~\cite{PDH} we only describe here its implementation in our setup.
An electro-optics modulator (EOM) has been placed at the exit of the laser oscillator (see figure~\ref{fig:schema_optique}). This EOM is used to modulate the frequency of the laser pulses by $\pm 5 MHz$. A photodiode is placed so as to see the part of the laser pulse that is reflected from the injection mirror M1. The signal from this photodiode is digitized by an ADC and then numerically demodulated in the FPGA (see figure~\ref{fig:elec_DPDH_plot}). This signal (input frequency; IF) is then feed to a  Finite Impulse Response filter (FIR), which has a band-pass frequency response.
 If the error is not too big the measured demodulated and filtered signal is directly proportional to the error on the injection time of the pulses in the FPC (see figure \ref{fig:elec_DPDH_plot}), however there  is not enough bandwidth to act directly on the length of the FPC.
 Instead the error signal is used to drive a PZT located inside the laser oscillator. 
 By acting on this PZT the length of the laser oscillator is changed so that the laser pulses are injected in the FPC with the proper timing and phase (see figure~\ref{fig:elec_full_system}). As the PZT has a limited range the length of the laser oscillator can also be changed by larger steps by adjusting its temperature. This allows us to reach any frequency within a 8kHz range.
To avoid changing the laser oscillator when no power is stored in the cavity and therefore the PDH error signal is not reliable a threshold is set on the minimum power that must be read by the photodiode located behind mirror M1 to allow actions on the laser oscillator PZT.

\begin{figure}
\begin{center}
\includegraphics[width=1.05\textwidth]{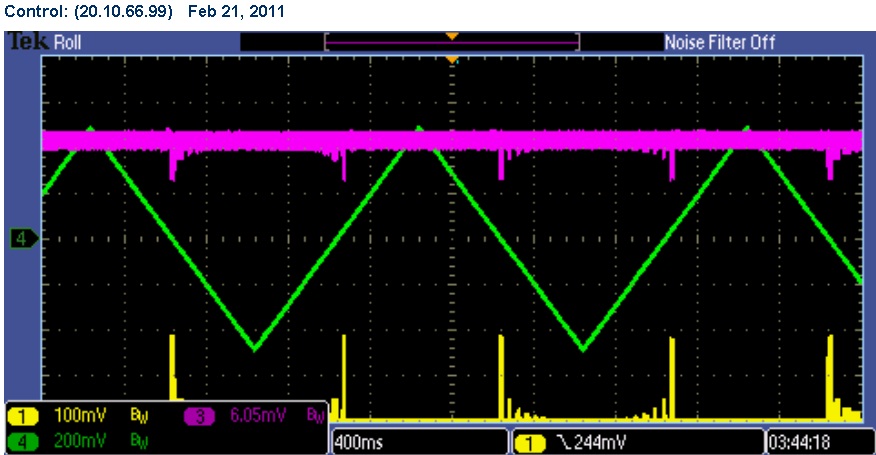} 
\includegraphics[width=1.05\textwidth]{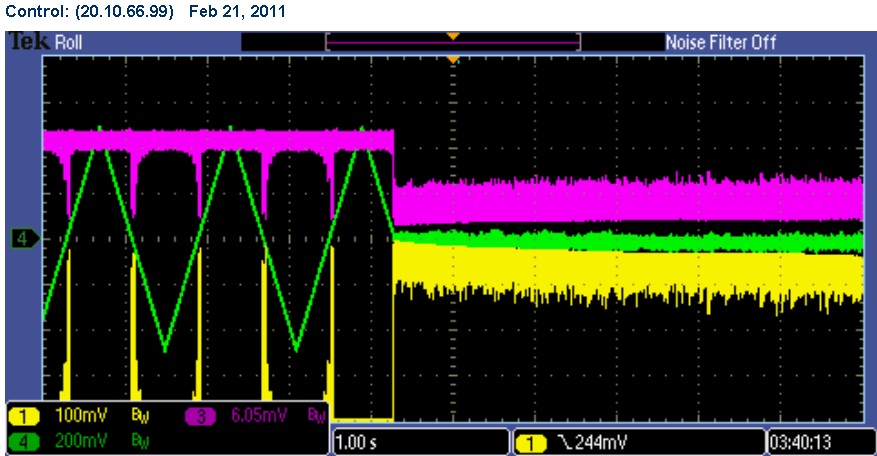} 
\caption{Oscilloscope traces showing the voltage applied to the PZT in the laser oscillator cavity (green), the power transmitted by the cavity (yellow) and the power reflected at the entrance of the cavity (magenta). 
To find the laser oscillator cavity length matching the FPC length we apply a triangular voltage on the PZT located in the laser oscillator cavity (upper plot). When the length of the laser oscillator becomes close a value that matches a mode in the FPC we see more power transmitted by the FPC (yellow) and less power reflected (magenta). If the power stored in the cavity becomes high enough to generate a usable PDH error signal the system automatically switches to that signal to adjust the laser oscillator cavity length (bottom plot). }
\label{fig:PZT_triangles}
\end{center}
\end{figure}

When no power is stored in the FPC the laser oscillator cavity length can be scanned by applying varying tension to the laser oscillator PZT. The range of this tension is adjusted manually around the values where we expect to see power accumulation in the FPC. An example of such scan is shown on figure~\ref{fig:PZT_triangles} (top).
 Once the PZT scans the range where power accumulation occurs sufficiently slowly for the PDH error signal to be meaningfull the system automatically switches to "locked" mode and uses the PDH error signal to vary the length of the laser oscillator as required to stay synchronized with the FPC. This is shown on the lower part of figure~\ref{fig:PZT_triangles}.

Although a great care has been taken while designing the feedback and selecting the components used to build it, 
it is necessary to construct a feedback filter which match the system's frequency response.
It is therefore necessary to measure how the whole system (device and feedback) reacts
to a given input and adjust it accordingly. This is done using an identification procedure~\cite {chapter_identification_gevers}.


The linear response of the system $y(z)$ can be described as follow~\cite{identification_Esmaili}:

\begin{equation}
y(z) = G(z)u(z) + H(z)e(z) 
\end{equation}

\begin{equation}
u(z) = K(z) [r(z) - y(z)]
\end{equation}
where z is the backward shift operator, $u(z)$ is the control input, $e(z)$ is white noise and $G(z)$ and $H(z)$ are the process transfer function, $K(z)$is the feedback transfer function and $r(z)$ is the reference signal of the feedback.

\begin{figure}
\begin{center}
\includegraphics[width=.8\textwidth]{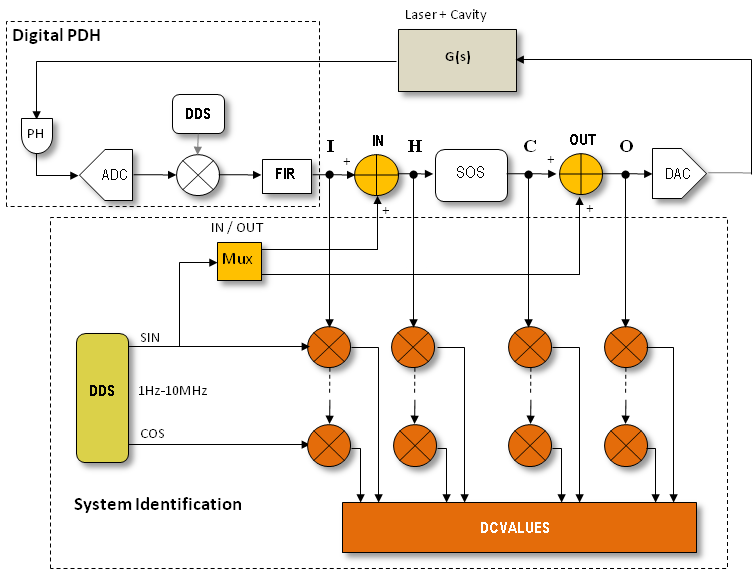} 
\caption{Block diagram of the laser oscillator to FPC feedback loop. To measure the response function of this feedback loop signal generated by the direct digital synthetizer (DDS) is injected in the loop before or after the SOS (locations "IN" or "OUT"). The response of the loop to this signal is read at locations I, H, C and O. The phase of this response is obtained by mixing it with the DDS signal. CHECK}
\label{fig:identification}
\end{center}
\end{figure}

Identification consists in estimating the parameters $G$ and $K$ 
by adding a known harmonic signal to the locked system and by making ratios between measured harmonic signals (see figure~\ref{fig:identification}).




Practically this is done by  injecting a known signal at various locations in our feedback loop  and measuring the response of the various elements of the system.
 This has been fully implemented in the FPGA which is to our knowledge the first time it was done for such system.
 
  Once the response of the system  has been measured it is possible to adjust the digital filters in the FPGA to bring its response as close a possible for the desired response and to minimize the impact of resonances in the feedback loop.


\begin{figure}[htbp]
\begin{center}
\includegraphics[width=.75\textwidth]{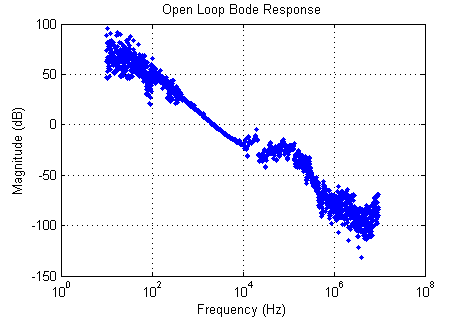} 
\caption{Bode diagram obtained by using the identification procedure to study the open loop response of our system. It can be seen that the system's response is unstable above 20~kHz. }
\label{fig:bode}
\end{center}
\end{figure}

\begin{figure}
\begin{center}
\hspace*{-2cm}\includegraphics[width=1.2\textwidth]{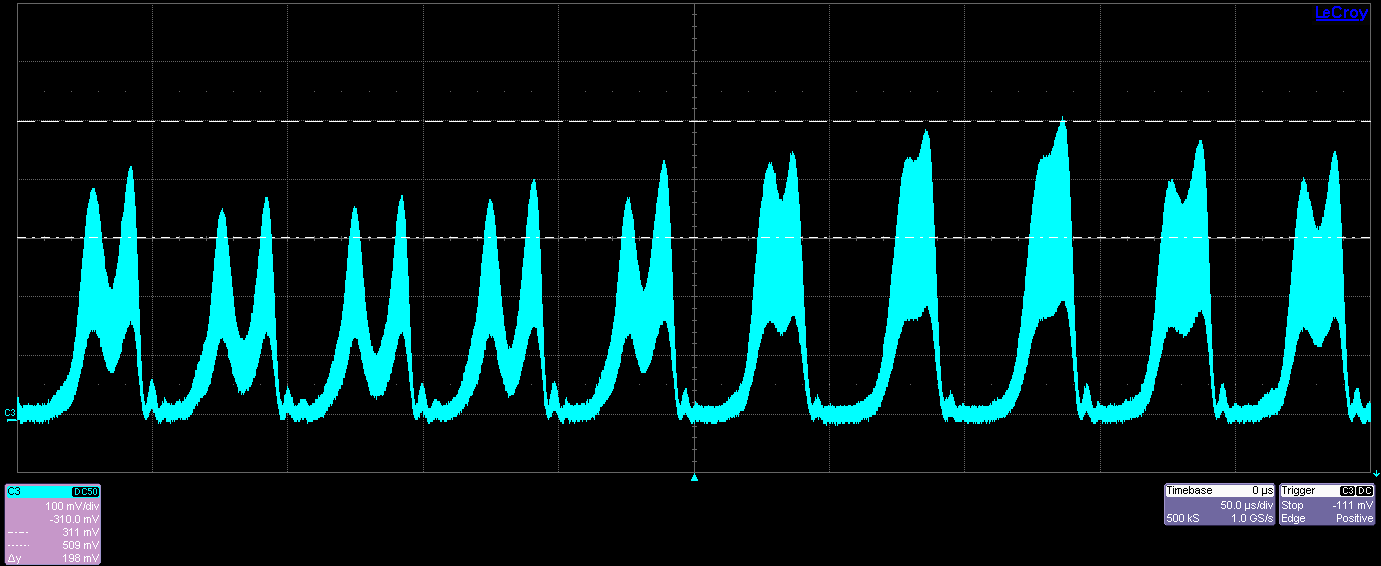}\\
\vspace*{0.2cm}
\hspace*{-2cm}\includegraphics[width=1.2\textwidth]{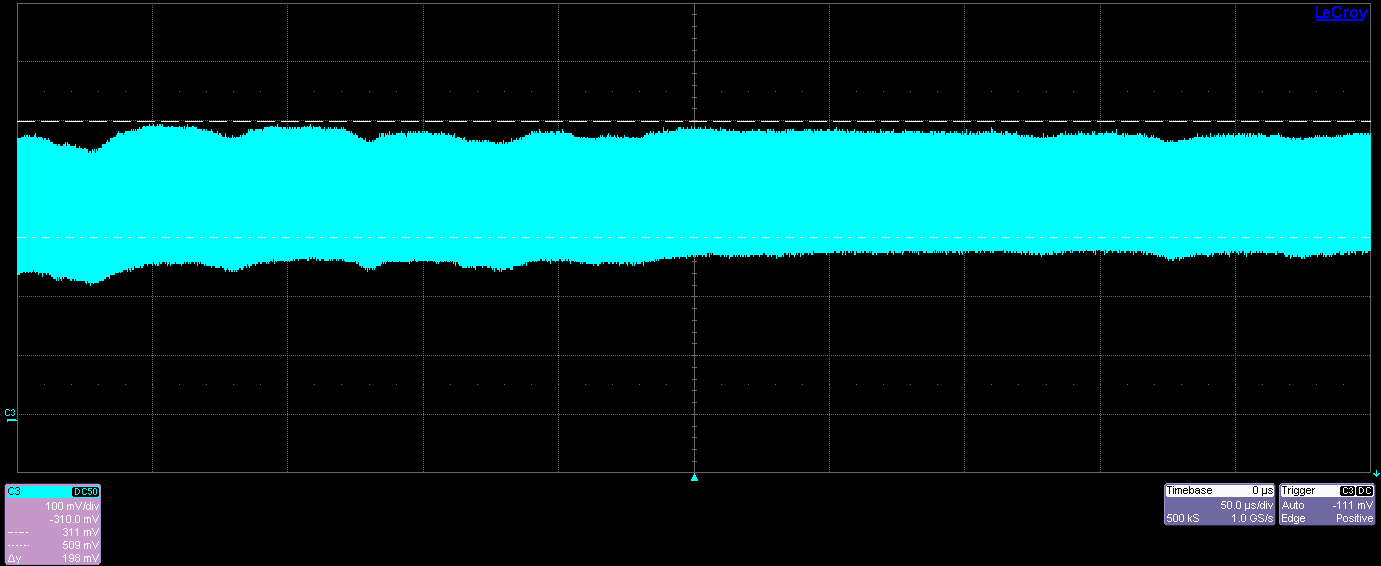}
\caption{Example of power stored in the FPC before (top) and after (bottom) adjusting the filters. On the top plot resonances at  20~kHz and 40~kHz can clearly be seen. These resonances disappeared after the filters were adjusted. The power stored in the cavity is measured by placing a photodiode behind one of the mirrors of the FPC (in our case mirror M2).}
\label{fig:scope_traces_identification}
\end{center}
\end{figure}

Figure~\ref{fig:bode} shows the open loop Bode diagram obtained after performing such measurement. On this diagram the 
system's response has peaks beyond 10 kHz.
This could also be seen by looking directly at the power stored in the cavity on a scope  (see figure~\ref{fig:scope_traces_identification} top). Using the knowledge gained by artificially stimulating our feedback system we modified the FPGA filters. After these modifications the power stored in the FPC was free of these resonances (see figure~\ref{fig:scope_traces_identification} bottom).

\section{Performances and future developments}

In this article, we described in details an apparatus, based on a non planar optical resonator, installed on the ATF electron ring. The system is meant to produce a circularely polarised high gamma flux from laser/ electron beam Compton scattering. This is the first time ever that a non-planar four mirrors cavity operating in pulsed mode and has been successfully deployed at an accelerator.

\begin{figure}
\begin{center}
\hspace*{-2cm}\includegraphics[width=1.2\textwidth]{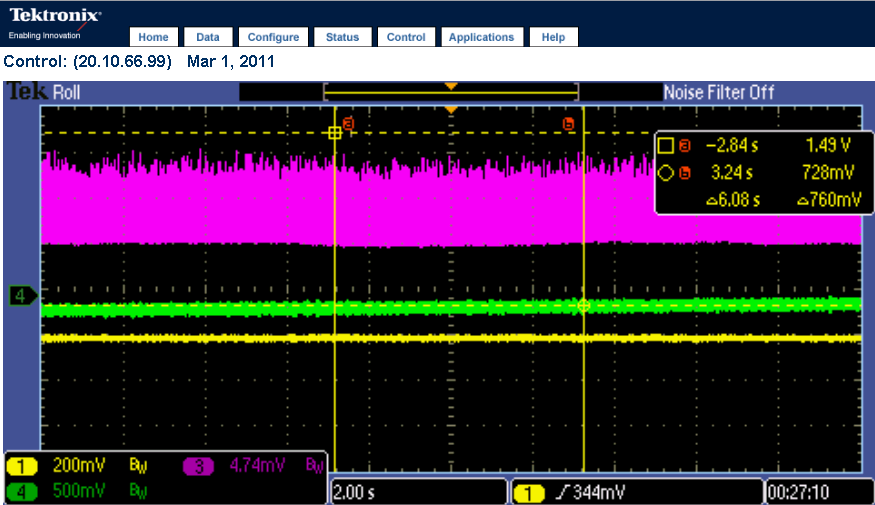} 
\caption{Oscilloscope trace showing the voltage applied to the PZT in the laser oscillator cavity (green), the power transmitted by the cavity (yellow) and the power reflected at the entrance of the cavity (magenta).  This trace shows the cavity locked during more than 20 seconds with a high level of stability. }
\label{fig:trace_locked}
\end{center}
\end{figure}

This apparatus has been installed at the KEK ATF during the summer 2010. The choice of a digital feedback made it easy to adapt it to the new system.
Within weeks of installing the cavity we were able to perform several  data taking run during which the two locks were kept during several hours (see figure~\ref{fig:trace_locked}). This has been validated experimentally by the observation of high energy gamma rays produced as soon as both the laser and the accelerator were operating. A flux of  2.7~$\pm$~0.2~gamma rays per bunch crossing ($\sim3\times10^6$~gammas per second) was achieved. The measurement of this flux and the analysis of the data taken is discussed in a companion paper~\cite{analysis_Iryna}. 

For the commissioning, the cavity finesse was fixed to a moderate value of $\approx 3000$. A dedicated fiber based amplifier allowed us to reach $\approx 55$W of incident laser beam average power. The best laser coupling to the cavity was at most at 65$\%$.

Before the earthquake that struck Japan in March 2011 we were working on several improvements to this system. To increase the power stored in the cavity, we intend to replace the current mirrors with higher reflectivity ones in order to achieve a finesse of 30~000. We also intend to improve the laser system and the injection optics to increase the power injected in the cavity. The great adaptability of the digital double feedback system gives us confidence that after increasing the cavity finesse we should still be able to lock.

The technology described in this article will be used to reach a final goal of 1MW average power stored inside a FPC with several possible applications.

\section*{Acknowledgements}

This work has been funded thanks to a grant from the French ANR (Agence Nationale de la Recherche) under contract number BLAN08-1-382932  and P2I (Physique des 2 Infinis). The authors would like to thank the staff from KEK for their hospitality and their help with the design, the installation and the operation of this FPC. The authors are also grateful to Franc{c}ois Richard for his continuous support.

\bibliographystyle{unsrt}
\bibliography{biblio}

\end{document}